\documentclass[prd,nofootinbib]{revtex4}
\usepackage{epsfig,amsmath}

\def\gs{{\gamma}_{S}}
\def\Qi{{\bf Q}_i}

\def\Qz{{\bf Q}}
\def\Q1{{\bf Q}_1}
\def\QN{{\bf Q}_N}
\def\Qis{{\bf Q}_{i;{\rm s}}}

\def\phiv{\mbox{\boldmath$\phi$}}

\def\ls{{\lambda_{S}}}

\def\i{{\rm i}}

\def\ee{e$^+$e$^-\;$}

\def\ssb{\langle {\rm s}\bar{\rm s}\rangle}
\def\ppb{${\rm p}\bar{\rm p}\;$}

\begin{document}


\title{Strange quark production in a
statistical effective model}
\author{F. Becattini}
\author{G. Pettini}
\affiliation{Department of Physics, University of Florence and INFN
Sezione di Firenze \\
Via G. Sansone 1, I-50109, Sesto F.no, Firenze, Italy}
\email{becattini@fi.infn.it, pettini@fi.infn.it}

\begin{abstract}                
An effective model with constituent quarks as fundamental degrees
of freedom is used to predict the relative strangeness production
pattern in both high energy elementary and heavy ion collisions.
The basic picture is that of the statistical hadronization model,
with hadronizing color-singlet clusters assumed to be at full
chemical equilibrium at constituent quark level. Thus, by assuming
that at least the ratio between strange and non-strange
constituent quarks survives in the final hadrons, the apparent
undersaturation of strange particle phase space observed in the
data can be accounted for. In this framework, the enhancement of
relative strangeness production in heavy ion collisions in
comparison with elementary collisions is mainly owing to the
excess of initial non-strange matter over antimatter and the
so-called canonical suppression, namely the constraint of exact
color and flavor conservation over small volumes.
\end{abstract}

\maketitle

\section{INTRODUCTION}
\label{int}

The recent observation that hadron multiplicities in \ee and
hadronic high energy collisions agree very well with a
statistical- thermodynamical ansatz ~\cite{beca1,beca2} has
revived the interest on the statistical models in high energy
collisions, an idea dating back to the 50's ~\cite{fermi,hagedorn}. As
the apparent chemical equilibrium of all hadron species in these
elementary collisions (EC) cannot be driven by inelastic
collisions amongst hadrons after their formation, this finding has
led to the idea of a pure statistical filling of multi-hadronic phase
space of homogeneous hadronizing regions ({\em clusters} or {\em fireballs})
as an intrinsic feature of
hadronization process, occurring at a critical value of energy
density~\cite{beca2,heinz,stock}. Otherwise stated, hadrons are
born in an equilibrium state, as envisaged by Hagedorn
\cite{hagedorn2}. It must be emphasized that, within this
framework, temperature and other thermodynamical quantities have
an essential statistical meaning which does not imply the
occurrence of a thermalization process via inelastic hadronic
collisions.

The same model, in different versions, has been successfully
applied to a large set of heavy ion collisions (HIC) data
\cite{hions,becahions2,becahions} and this has been interpreted as
a clue of the minor effect of post-hadronization inelastic
rescattering \cite{heinz}, an indication which is also supported
by kinetic models calculations \cite{kinetic}.

One of the nice features of the statistical hadronization model
(SHM) is the very low number of free parameters required to
reproduce a large number of hadronic multiplicities. Provided that
the masses and charges of the assumed hadronizing clusters
fluctuate according to a particular shape function
\cite{beca2,becapt}, all Lorentz-invariant quantities (like e.g.
hadron average multiplicities) depend on two parameters: the sum
$V$ of the volumes of the clusters and the
temperature $T$. However, in order to get a satisfactory agreement
with the data, the model has to be supplemented, both in EC and
HIC, with one more phenomenological parameter, $\gs$, suppressing
the production of particles containing $n$ strange valence quarks
by $\gs^n$ with respect to the expected production in a fully
equilibrated hadron gas \footnote {Recently, a different
parameterization of the extra strangeness suppression has been
introduced for elementary collisions \cite{becapt}.}.

\begin{figure}
\caption{$\ls$ in various elementary and heavy ion collisions
(from \cite{becasqm}).
\label{lambdas}}
\includegraphics{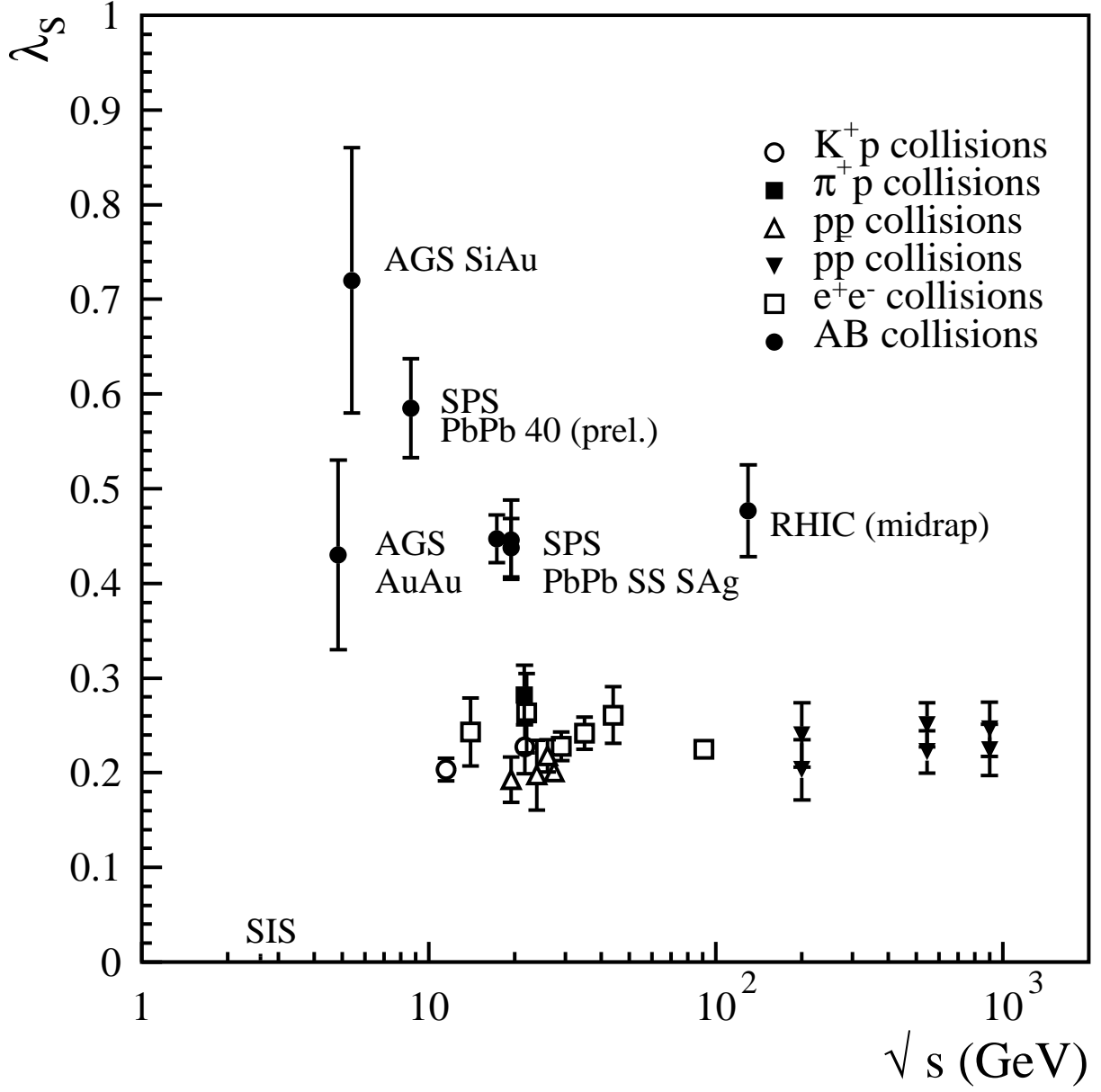}
\end{figure}

The behavior of $\gs$ as a function of collision and
center-of-mass energy was found to be rather odd as it turns out
to be significantly higher in \ee collisions than in hadronic
collisions over a large energy range \cite{beca1,beca2} whereas it
has a fairly stable value in heavy ion collisions
\cite{becahions2}, which is approximately the same as in \ee
collisions. A clearer insight in the mechanism of strangeness
production can be achieved by estimating the ratio between newly
produced valence strange and up, down quark pairs, the so-called
Wroblewski factor $\ls$:
\begin{equation}
  \lambda_{S} = \frac{2 \langle {\rm s \bar s} \rangle}
   {\langle {\rm u \bar u} \rangle + \langle {\rm d \bar d} \rangle}
\label{landas}
\end{equation}
which shows a striking regularity, being in fact fairly constant
in all kinds of elementary collisions \cite{becahions} over two
orders of magnitude of center-of-mass energy and as twice as large
in high energy heavy ion collisions (see Fig.~\ref{lambdas}). It
should be stressed that this ratio is calculated by counting the
{\em primarily} produced quark pairs, i.e. those belonging to
directly emitted hadrons. As primary hadrons are not measurable,
they must be calculated by using a model and $\ls$ turns out to be
a model dependent quantity especially with regard to the number of
u and d quarks which significantly increases during the
post-hadronization hadronic decay chain. However, if the number of
measured species is large and the model accurately reproduces them
(which is the case for the SHM), the thereby estimated $\ls$ is
expected to be reasonably close to the actual value. The need of
an extra suppression of strangeness with respect to the full
statistical equilibrium for the hadronic system urges the search
for an explanation based on a more microscopic approach. This is
indeed the main subject of the present work, where we try to
calculate $\ls$ within one statistical model scheme in both EC and
HIC, employing a constituent quark model as the basic underlying
structure, with the purpose of justifying the lack of strangeness
chemical equilibrium at hadron level observed in the data. In
fact, constituent quark models have already been used to calculate
strangeness production in HIC on the basis of transport equation
\cite{aichreb}. Amongst other microscopic approaches to this
problem and, more generally, to account for hadronization
equilibrium features, a recent study has been performed in
ref.~\cite{dumitru} focussed on the role of massive Polyakov
loops.

The paper is organized as follows: in Sect.~\ref{sec:physics} we
introduce the physical picture and in Sect.~\ref{sec:model} the
full model is described. In Sect.~\ref{sec:appres} the
Nambu-Jona-Lasinio model with exact conservation of quantum
charges is used to calculate $\lambda_{S}$ and compare it with the
corresponding value obtained from fits of the statistical
hadronization model to the data in both HIC and EC. Also, the
stability of the results is addressed.
Sect.~\ref{sec:conclu} is devoted to conclusions.

\section{THE PHYSICAL PICTURE}
\label{sec:physics}

The physical picture keeps the same scheme of the SHM: formation
of a set of clusters endowed with charge, momentum, mass and
volume, in local statistical equilibrium. However, statistical
equilibrium is now assumed to apply to a system of constituent
quarks whilst hadrons are assumed to be produced via their
coalescence, still in a purely statistical fashion ({\em
statistical coalescence}) so as to give rise to a {\em partially}
chemically equilibrated hadron gas. Furthermore, each cluster is
required to be a color singlet. This requirement has of course no
effect at hadron level, but it is crucial in a quark model.

Two more crucial assumptions are introduced. First, the thermodynamical
parameters fitted in the SHM, in high energy EC and HIC, are interpreted as
the critical values for color deconfinement and, consequently, for
(approximate) chiral symmetry restoration \cite{heuristic,karsch};
secondly, it is assumed that the produced s and light quarks, or
at least the ratio s/u, i.e. $\ls$, survive in the final hadrons
after hadronization has taken place.

The first assumption is supported by the constancy of the fitted
temperatures in the SHM for various processes in EC \cite{becabiele} and
by its agreement with other estimates of the critical temperature;
in view of this fact, the use of effective models such as the
Nambu-Jona-Lasinio model \cite{njlvero,kunihiro} and others
\cite{bcd,tricritical,conmassa,giap} embodying chiral symmetry
breaking and restoration, looks well suited. It should be pointed
out that the values of the thermodynamical parameters (i.e. temperature
and baryon-chemical potentials) extracted within the SHM in HIC
(see Table \ref{tabbeca}) lie in a range where all effective models
and recent lattice simulations \cite{fodor} predict smooth
cross-over transitions,
far from a possible critical ending point. For the same reason,
extensions of the NJL model \cite{klevalast,aichelin} at high
baryon chemical potential and low temperature, where color
superconductivity takes place\cite{cfl}, can be disregarded.

The second assumption allows us to calculate $\ls$ within a quark
model and compare it with that obtained from valence quark counting
in the produced hadrons. Also, it implies that the lack of full chemical
equilibrium at hadron level is indeed the consequence of full
chemical equilibrium at constituent quark level.

In the grand-canonical framework, which applies to a large system,
$\ls$ would depend only on intensive quantities such as chemical
potentials and temperature and its constancy would be tightly
related to theirs. However, if the system is not large, the
canonical (and possibly micro-canonical) ensemble, in which the
exact conservation of cluster quantum numbers in the hadronic or
quark system is enforced, must be used and, as a consequence,
$\ls$ gets a dependence also on the volume (the so-called {\em
canonical suppression}). While the volume within which quantum
charges (baryon number, electric charge and strangeness) are fixed
can be taken as the sum of the proper volumes of all clusters,
though under appropriate non-trivial assumptions
\cite{beca2,becapt}, the volume over which color must be
neutralized is, as has been mentioned, that of a single
hadronizing cluster. This very fact introduces a further parameter
in the model, i.e. the average single cluster volume $V_c$ which
can indeed be much smaller than the volume over which flavor is
conserved. As quarks do carry color charge, $\ls$ might be
significantly affected by variations of $V_c$.

Therefore, the underlying idea is that the presence of a
characteristic constant value of $V_c$ in EC, related to the typical
distance over which color is neutralized, may sizeably reduce the value
of $\ls$ (together with flavor constraint) whereas color deconfinement
in HIC may lead to an enhancement of s/u ratio. We thus argue that the
main difference between hadronization in EC and HIC is to be found in
the typical size of the pre-hadronization color-neutral region: it has
to be something of the order of a hadron radius in EC (a sort of
mini-QGP droplet) and much larger in HIC if a macroscopic (i.e.
extending over many hadron volumes) QGP has been formed. Of
course, it must be emphasized that the process stage at which the
statistical equilibrium is achieved is expected to be deeply
different in EC as compared with HIC: an {\em early}
thermalization of partons is indeed envisaged in HIC, which is
maintained until hadronization, whereas a {\em late},
pre-hadronization local equilibrium scenario, possibly driven by
the strongly coupled non-linear evolution of QCD fields in the
late soft regime is envisaged for EC.

Hence, we argue that relative strange quark production might probe
these two scenarios because of a possible stronger color and
flavor-canonical suppression of s quarks with respect to u and d
quarks in EC which is absent in HIC due to much larger single-cluster
volume and overall volume.
The ultimate reason of the stronger canonical suppression of s quarks
in comparison with u, d, resides in their different constituent mass
values. The hereby proposed mechanism of strangeness
enhancement in HIC is rather different from that suggested by
Muller and Rafelski \cite{rafe} insofar as no difference in the time
scale is invoked but an equilibrium situation is assumed in both
EC and HIC, though over differently sized regions.

\section{THE MODEL}
\label{sec:model}

According to the basic idea of the SHM and what has been said in
the previous section, the formation of a set of pre-hadronic,
color-singlet clusters at statistical equilibrium, with definite
values of flavor quantum numbers is envisaged as the result of a
complex dynamical evolution. In an equilibrium scheme, all
physical observables relevant to a particular cluster can be
calculated by means of suitable operations on its associated
partition function. Provided that cluster masses are large enough
so that micro-canonical effects can be neglected, the requirement
of definite quantum numbers implies that the $i^{\rm th}$
cluster's partition function has to be the {\em canonical} one
rather than the more familiar grand-canonical:
\begin{equation}
  Z_{i} = \sum_{\rm states} e^{-E/T}
\end{equation}
where $T$ is the cluster's temperature, $E$ its energy and the sum
is meant to run over all {\em allowed} states, namely the states
with flavor and color quantum numbers matching those of the
cluster.
The calculation of $Z_{i}$ can be done by using the symmetry group
$G$ associated to the involved quantum numbers. If the {\em
continuous} group $G$ is an exact internal symmetry for the
Hamiltonian $H$ and the cluster state belongs to its irreducible
representation $\nu$, then \cite{exact}:
\begin{equation}\label{zcan}
 Z_{i} = \int d\mu(\gamma_{1},...,\gamma_{r})
 \chi^{*}_{\nu}(\gamma_{1},...,\gamma_{r})
 {\hat Z}_{i}(\gamma_{1},...,\gamma_{r}) =
 \int d\mu(\gamma_{1},...,\gamma_{r}) \chi^{*}_{\nu}(\gamma_{1},...,\gamma_{r})
 {\rm Tr}~\Big(e^{\displaystyle {-\beta H - \i \sum_{l=1}^{r}\gamma_{l} Q_{l}}}\Big)
\end{equation}
where $d\mu(\gamma_{1},...,\gamma_{r})$ is the invariant
normalized measure of $G$, $\chi_{\nu}$ is the character of the
representation $\nu$, $r$ is the rank of $G$, $Q_{l}$ are the
mutually commutating generators of the Cartan subalgebra and
$\gamma_{l}$ are the group parameters. The actual group of
interest is SU(3)$\times$U(1)$^3$, where each U(1) corresponds to
conserved net flavors $U$, $D$ and $S$ and SU(3) to the color. For
instance, the canonical partition function of the $i^{\rm th}$
cluster of a gas of free quarks (no antiquarks included) with
flavor $j$ involves the following $\hat Z_i$ in Eq.~(\ref{zcan}):
\begin{equation}\label{fermions}
  {\hat Z}_{i}(\theta_{1},\theta_{2},\phi_{j})=
  \exp \left(\sum_{n=1}^{\infty}
   {(-1)^{n+1}\over n} \chi_{1,0}(n\theta_{1},n\theta_{2})\,
   e^{\i\,n\phi_{j}}~\frac{g V}{(2\pi)^3} ~
   \int d^{3}p~e^{\displaystyle{-{n\epsilon_{j}\over T}}}\right)
\end{equation}
where:
\begin{equation}
 \chi_{1,0}(\theta_{1},\theta_{2})=
 e^{\displaystyle{-\i \theta_{1}}}+e^{\displaystyle{-\i\theta_{2}}}
 + e^{\displaystyle{\i\left(\theta_{1}+\theta_{2}\right)}}
 \label{caracara}
\end{equation}
is the character of the fundamental representation of SU(3), with
$\theta_{1,2} \in [-\pi,\pi]$; $\phi_{j}$ is the parameter of U(1)$_j$
$V$ is the volume, $g$ is the spin degeneracy factor, $\epsilon_{j}=
\sqrt{p^2+m_{j}^{2}}$ and the sum over discrete states $k$ has been
approximated with its continuum limit. When including antiquarks and
considering three flavors (i.e. the full group
SU(3)$\times$U$_{u}$(1)$\times$U$_{d}$(1)$\times$U$_{s}$(1)),
${\hat Z}^{i}$ can be written as the product of three functions
like those in Eq.~(\ref{fermions}):
\begin{equation}\label{totalhatz}
 {\hat Z}_{i}\left(\theta_{1},\theta_{2},\phi_{u},\phi_{d},\phi_{s}\right)
  =\prod_{j=u,d,s}{\hat Z}_{ij}\left(\theta_{1},\theta_{2},\phi_{j}\right)
  ~{\hat Z}_{ij}^*\left(\theta_{1},\theta_{2},\phi_{j}\right)
\end{equation}
The irreducible representation $\nu$ can be labelled by three integer
numbers, the net flavor $U_i,D_i,S_i$, and, as far as the SU(3)
color group is concerned, $\nu$ is simply the singlet representation.
Therefore, by introducing the vectors $\Qi = (U_i, D_i, S_i)$ and
$\phiv_i = (\phi_{iU}, \phi_{iD},\phi_{iS})$, the character can be
written as $\chi_{\nu}=\exp[-\i \Qi \cdot \phiv_i]$ and:
\begin{equation}\label{zeta}
  Z_{ij} = \left[ \prod_{j=1}^3 \int_{-\pi}^{\pi} {d\phi_j \over 2\pi}
  \right] \int d\mu(\theta_{1},\theta_{2}) e^{\displaystyle{\i \Qi \cdot \phiv_i}}
 {\hat Z}_{i}\left(\theta_{1},\theta_{2},\phiv_i \right)
\end{equation}
where:
\begin{equation}
\int d\mu (\theta_{1},\theta_{2})={1\over 3!}
\int_{-\pi}^{\pi}{d\theta_{1}\over 2\pi} \int_{-\pi}^{\pi}
{d\theta_{2}\over 2\pi} \prod_{j<k}^{3} 4 \sin^2{{\theta_{j}-\theta_{k}}
\over 2}
\label{sutremis}
\end{equation}
with the constraint $\sum_{k=1}^{3}\theta_{k}=0 ~{\rm mod}~2\pi$,
is the SU(3) invariant integration. The partition function
hitherto considered is relevant to one cluster. Yet, there are
several clusters in a single collision event and, as long as global
observables are concerned, theoretical predictions require summation
over all clusters. Moreover, clusters may well be produced with different
configurations of flavor numbers $(U_i,D_i,S_i)$ in different
events (provided that their sum fulfills the conservation law,
i.e. it must be equal to the flavor numbers of the colliding
particles), hence an integration over all possible configurations
weighted by its probability must be carried out in order to
calculate averages of physical quantities. However, the
probability distribution of clusters flavor configuration is an
unknown function, which cannot be predicted within the statistical
model and which is most likely governed by the preceding dynamical
process. We will therefore assume a configuration probability
distribution $w$ which is the ``maximum disorder" one and which
allows a remarkable simplification of the expressions of averages
quark multiplicities as well as any other Lorentz-invariant
observables. This function $w$, for an event with $N$ clusters can
be written as:
\begin{equation}\label{w}
  w(\Q1,\ldots,\QN) = \frac{\prod_{i=1}^N Z_{i}(\Qi)
  \delta_{\Qz,\Sigma_i \Qi}}{\sum_{\Q1,\ldots,\QN} \prod_{i=1}^N
  Z_{i}(\Qi) \delta_{\Qz,\Sigma_i \Qi}}
\end{equation}
where $Z_i$ as in Eq.~(\ref{zeta}), the $\delta$ ensures that
global conservation of initial quantum numbers $\Qz$ is fulfilled.
If one has to calculate the average multiplicity of a given quark
species $j$ in a $N$ cluster event, it is advantageous to introduce
a fictitious fugacity $\lambda_j$ and taking the derivative of
$\log Z_i$ with respect to it, so that the overall multiplicity
reads:
\begin{eqnarray}\label{mult}
 \langle n_{j} \rangle &=& \sum_{\Q1,\ldots,\QN} w(\Q1,\ldots,\QN)
 \sum_{i=1}^N \frac{\partial}{\partial \lambda_{j}} \log Z_{i}
 (\Qi)\Big|_{\lambda_{j}=1} = \nonumber \\
 &=& \sum_{\Q1,\ldots,\QN} w(\Q1,\ldots,\QN) \,\,
 \frac{\partial}{\partial \lambda_{j}} \prod_{i=1}^N \log Z_{i}
 (\Qi)\Big|_{\lambda_{j}=1}
\end{eqnarray}
and, by using Eq.~(\ref{w}):
\begin{equation}\label{mult2}
 \langle n_{j}\rangle = \frac{\partial}{\partial \lambda_{j}}
 \log \sum_{\Q1,\ldots,\QN} \prod_{i=1}^N Z_i(\Qi) \,
 \delta_{\Qz,\Sigma_i \Qi} \Big|_{\lambda_{j}=1}
\end{equation}
Were not for the color-singlet constraint stuck to each $Z_i$, the
sum over all configurations on the right hand side of
Eq.~(\ref{mult2}) would be, for $\lambda_{j}=1$, the canonical
partition function of a single cluster having as volume the sum of
volumes of the individual clusters and with quantum number vector
$\Qz~$\cite{beca2}, provided that all clusters have the same
temperature. Thereby, also the explicit dependence on $N$ would
vanish. However, this is no longer true once the color singlet
constraint is introduced. For a similar strong equivalence with
one cluster to apply, single clusters should be in suitable color
quantum states superpositions, which is definitely against the
common belief of pre-confinement and local color neutralization.
Nevertheless, it can be proved that a considerable simplification
in Eq.~(\ref{mult2}) occurs if clusters in one event are assumed
to have the same volume, i.e. $V_i \equiv V_c$. In that case, it
can be proved (see Appendix A) that the right hand side of
Eq.~(\ref{mult2}), which can be defined as the {\em global
partition function}, reads:
\begin{equation}\label{zetafinal}
  Z = \left[ \prod_{j=1}^3 \int_{-\pi}^{\pi} {d\phi_j \over 2\pi}
  \right] e^{\displaystyle{\i \Qz \cdot \phiv}}
  \left[ \int d\mu(\theta_{1},\theta_{2}) \;
  {\hat Z}_{i}\left(\theta_{1},\theta_{2},\phiv \right) \right]^{V/V_c}
\end{equation}
where $V=\sum_i V_i = N V_c$ (for the free quarks gas ${\hat
Z}_{i}$ is given by Eq.~(\ref{totalhatz}) with the single flavor
term ${\hat Z}_{ij}(\theta_1,\theta_2,\phi_j)$ calculated from
Eq.~(\ref{fermions}) with (\ref{caracara})). The introduction of
the local color-singlet constraint gets the value of the global
partition function severely reduced with respect to an only {\em
globally} color-singlet constrained case where clusters are
allowed to be in a quantum superposition of colored states. This
can be quite easily understood for any extra constraint
essentially involves a decrease of the number of available states
in the system. If $V_c \equiv V$ and $V$ is very large, i.e. in the
thermodynamical limit, it can be shown that the usual grand-canonical
expressions of the partition function, average multiplicities etc.
are recovered.

Besides the basic assumption of local statistical equilibrium (for
a single cluster), the expression~(\ref{zetafinal}) ultimately
relies on a particular choice of flavor distribution (i.e.~(\ref{w}))
among clusters, which is indeed a non-trivial assumption. This
particular distribution allows the reduction to
one equivalent global cluster just because it is the ``maximum
disorder" distribution, namely the probability of obtaining a
configuration $(\Q1,\ldots,\QN)$ by randomly splitting a
statistically equilibrated global cluster into $N$ sub-cluster
each with volume $V_c$ and in a color singlet state ; in fact,
this can be proved by showing that the $w$'s in Eq.~(\ref{w})
minimize the free energy of the system (see Appendix B).

It should pointed out that hadronizing clusters might be too small
for a local (i.e. in each of them) temperature to be defined, as
we have tacitly assumed. Should this be the case, the appropriate
treatment would be micro-canonical and not canonical, with
clusters described by mass and volume instead of temperature and
volume. Notwithstanding, as far as the calculation of
Lorentz-invariants is concerned, a global temperature may still be
recovered by resorting to a similar equivalent global cluster
reduction procedure at micro-canonical level, which is described
in detail in ref.~\cite{becapt} \footnote{Therein, the proof has
been given for a hadron gas but it can be easily extended to the
present case.}, with a micro-canonical equivalent of the
distribution~(\ref{w}). The proof is quite lengthy and the
mathematical framework more involved, but it is no longer
necessary to assume that each cluster ought to have the same
temperature. In fact, only the equivalent global cluster or,
equivalently, the global canonical partition function~(\ref{zetafinal}),
must be large enough to allow a canonical treatment of the system
as a whole.

We now have to face the problem of a suitable choice of a microscopic
Hamiltonian to be used in Eq.~(\ref{zcan}) and, thereby, make a theoretical
calculation of $\lambda_S$. As full QCD is out of question, we
resort to QCD-inspired low energy models. The simplest model to start
with is the Nambu-Jona-Lasinio (NJL) model, which has quarks as
fundamental degrees of freedom and embodies essential features
of dynamical chiral symmetry breaking ($\chi$SB) in the limit of
vanishing current quark masses. Also, we confine ourselves to
mean-field approximation, as presented in ref.~\cite{kunihiro},
using the effective Hamiltonian:
\begin{equation}\label{hnjl}
H^{^{NJL}}_{MFA}=V\left[g_{_{S}}(\alpha^2
+\beta^2+\gamma^2)+4g_{_{D}}\alpha\beta\gamma\right]~+~ \int d^3
x~{\bar q}\left( -\i{\bf \gamma}\cdot{\bf \nabla}+{\bf M}\right)q
\end{equation}
where $V$ is the spatial volume, $g_{_{S}}$ and $g_{_{D}}$ are the
four-fermion and the six-fermion $U(1)_{A}$-breaking couplings
respectively \cite{kunihiro,uauno}, $\alpha=\langle {\bar u}u\rangle,\beta=\langle {\bar
d}d\rangle,\gamma=\langle {\bar s}s\rangle$ are the quark
condensates and:
\begin{eqnarray}\label{constmas}
M_{u}&=&m_{u}-2g_{_{S}}\alpha-2g_{_{D}}\beta\gamma\nonumber\\
M_{d}&=&m_{d}-2g_{_{S}}\beta-2g_{_{D}}\alpha\gamma\nonumber\\
M_{s}&=&m_{s}-2g_{_{S}}\gamma-2g_{_{D}}\alpha\beta
\end{eqnarray}
are the constituent quark masses which, owing to the contact
four-fermion interaction, are linearly related to the current
quark masses $m_u,m_d,m_s$. As far as the grand-canonical
calculation is concerned, the variational parameters
$\alpha,\beta,\gamma$ are determined by minimizing the effective
potential \cite{kunihiro} obtained within standard methods
\cite{cjt,dolan}. Of course, the same result can be readily obtained
by calculating:
\begin{eqnarray}
Z&=&{\rm Tr}~e^{\displaystyle
{-\beta \Big(H-\sum_{i}\mu_{j}{\hat N}_{j}\Big)}}\nonumber\\
&=&e^{\displaystyle{-\Gamma |_{\rm min}}} \label{zetagr}
\end{eqnarray}
with $H=H^{^{NJL}}_{MFA}$. In Eq.~(\ref{zetagr}), ${\hat N}_{j}$
are the conserved flavor operators and $\Gamma|_{\rm min}$ is the
effective action at the physical point. For constant fields
$\Gamma|_{\rm min}=\beta V {\cal V}|_{\rm min}=-{\rm log}Z$, where
${\cal V}$ is the effective potential.\\
The quark over antiquark excess is obtained by taking the
derivative of ${\cal V}$ with respect to the chemical potential
for a given flavor:
\begin{equation}
\langle n_{j}-{\bar n}_{j}\rangle=-{\partial {\cal V}|_{{\rm min}}\over
\partial\mu_{j}} \label{defnum}
\end{equation}
whereas the number of particles, in the grand-canonical ensemble,
reads:
\begin{equation}
\langle n_{j}\rangle={N_{c}~V\over \pi^2}~\int_{0}^{\Lambda} dp
{p^{2}\over e^{\displaystyle{\beta\left(
\sqrt{p^2+M_{j}^2}-\mu_{j}\right)}}+1} \label{freenumb}
\end{equation}
where $\Lambda$ is an UV cutoff and $N_c$ the number of colors.

One of the remarkable features of the Hamiltonian $H^{^{NJL}}_{MFA}$
in Eq.~(\ref{hnjl}) is that the previously obtained expressions in
Eqs.~(\ref{zeta}), (\ref{zetafinal}) for the free quark gas partition
functions with exact flavor and color conservation, can be taken up
by simply replacing the current masses with the constituent
masses $m_{j}\rightarrow M_{j}$ and setting the cutoff $\Lambda$ as upper
bound for the momentum integration in Eq.~(\ref{fermions}).

The NJL model is the simplest choice in order to study strange
quark production, yet it is not a compelling one. Actually, its
validity is limited within a temperature range well below the UV
cutoff $\Lambda$ \cite{aichelin}. However, we are reasonably
confident that the main results found within this model would not
essentially change when employing other effective models whose
validity, in principle, extend to arbitrarily high temperatures.
As an example, it is worth mentioning the model developed in
refs.~\cite{bcd,tricritical,conmassa}, named in
ref.~\cite{kunihiro} as {\em ladder-QCD}. In this model the
exchange of a gauge particle between quarks is considered,
implying a momentum dependence of the self-energy and,
consequently, an irreducibility to a simple model with free
constituent quarks as in NJL. However, {\em ladder-QCD} is
considerably different from NJL only in the UV regime, where a
$1/p^2$ tail is added to the self-energy, which is otherwise
constant in the small momentum region\cite{bcd,russi}. The effect
of the different behaviour in the UV regime on the integrated
number of particles is expected to be rather small \cite{bcd}.

\section{ANALYSIS AND RESULTS}
\label{sec:appres}

\begin{figure}
\caption{Chiral susceptibility $\chi_{m}$ for the quark u or d
with current mass $5.5$ MeV as a function of temperature $T$, for
different values of the $T_{0}$ parameter in
the NJL model. The baryon chemical potential $\mu_{B}$ has been
set to 238 MeV and $\mu_Q$ has been set to 0, in accordance with
the analysis of full phase space Pb--Pb data at SPS energy
\cite{becahions2}. The maximum of $-\chi_{m}$ is located at $T =
158$ MeV, the fitted value in Pb-Pb collisions at SPS energy for
$T_{0}=170$ MeV.
\label{susc}}
\includegraphics{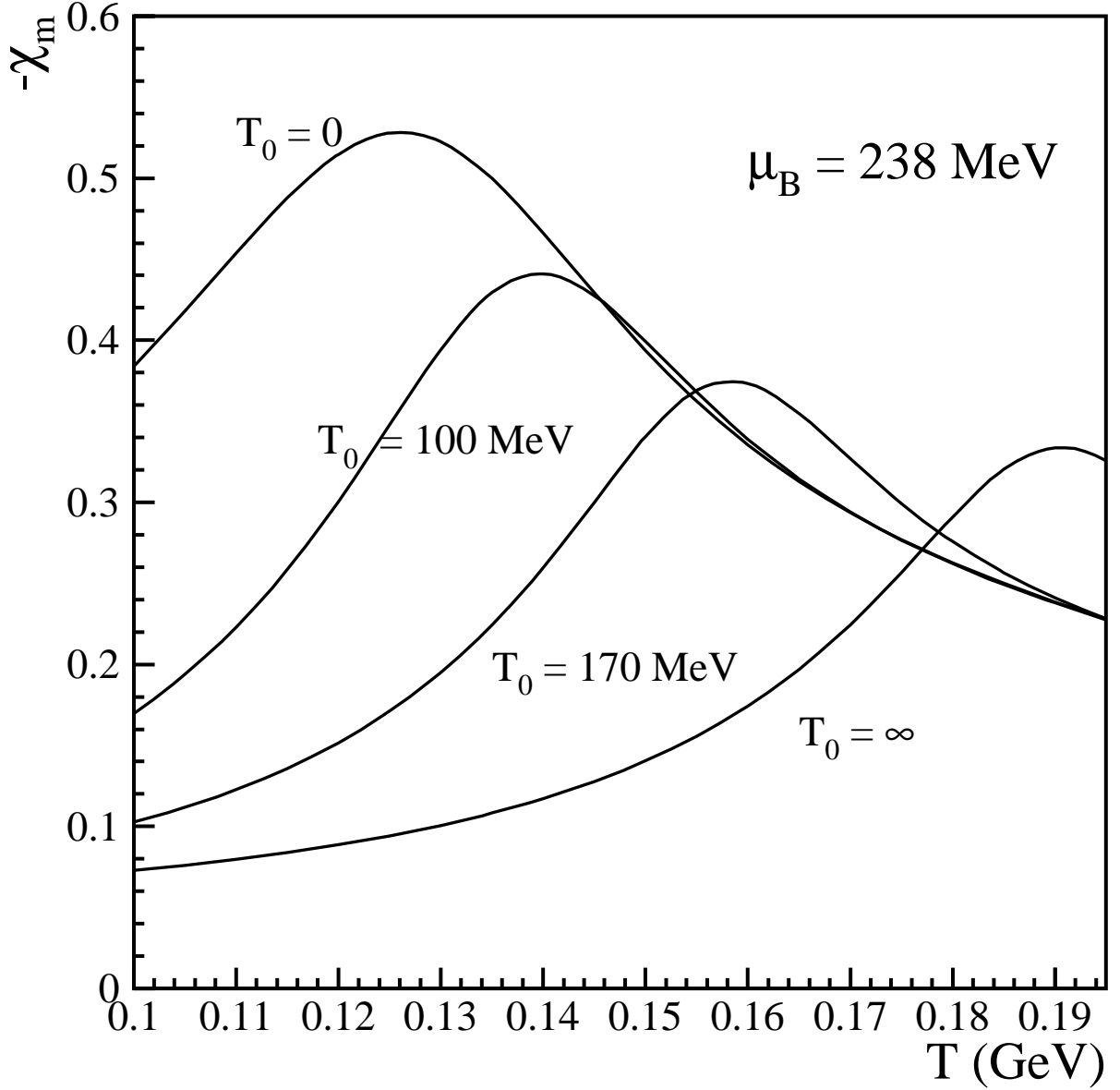}
\end{figure}

The goal of the numerical analysis is the calculation of $\lambda_{S}$
in both HIC and EC, by performing the ratio between newly produced
strange quark and u, d quark pairs in the NJL model. The main ingredient
needed to calculate $\lambda_S$ are the constituent quark masses which, for
a given temperature and baryon-chemical potential, have to be
determined by the minimization of the free energy $F$ and thus
depend on the parameters contained in the effective
Hamiltonian~(\ref{hnjl}), i.e. $g_S$, $\Lambda$, $g_D$ and the current
quark masses. Their values have been fitted at zero temperature and
baryon density in ref.~\cite{kunihiro} to many static and dynamical meson
properties:

\begin{eqnarray}\label{fitnjl}
{\hat m}&=&{m_{u}+m_{d}\over 2}=5.5~{\rm MeV};~~~~~~~~m_{s}=135.7
~{\rm MeV}\nonumber\\
\Lambda&=&631.4~{\rm MeV};~~~g_{_{S}}\Lambda^{2}=3.67;~~~
g_{_{D}}\Lambda^{5}=-9.29
\end{eqnarray}
Particularly for $g_D$, in ref.~\cite{kunihiro} a temperature dependence
is introduced through the phenomenological law:

\begin{equation}
g_{D}(T)=g_{D}(T=0)~e^{-\left(T/T_{0}\right)^2} \label{njlcases}
\end{equation}
where $T_0$ is a further free parameter and $g_{D}(T=0)$ is quoted
in Eq.~(\ref{fitnjl}). Different values of $T_0$ give rise to
different behaviors of the constituent u, d mass (see
Fig.~\ref{masses} later on) as a function of the temperature as
well as different positions of the cross-over curve in the
$(\mu_B,T)$ plane. According to the statement in
Sect.~\ref{sec:physics}, we take the thermodynamical parameters
$T$ and $\mu_B$ in Table \ref{tabbeca} as an input, and therefore
$T_0$ can be fixed by enforcing that the quark susceptibility for
a light quark with mass $\hat{m}$ in Eq.~(\ref{fitnjl}):
\begin{equation}
\chi_{m}={\partial\langle{\bar u}u\rangle\over
\partial m}\Bigg|_{m={\hat m}}
\label{suscept}
\end{equation}
has a peak at the desired $(\mu_B,T)$ point. This has been done
for the most accurately determined thermodynamical parameters,
which are those in Pb-Pb collisions at $\sqrt s_{NN} = 17.3$ GeV,
yielding $T_0 = 170$ MeV (see Fig.~\ref{susc}). Thereby, all
relevant parameters are now fixed. The only free parameter left is
the single cluster proper volume $V_{c}$, which may vary as a
function of center-of-mass energy and colliding system. The
dependence of $\lambda_S$ on this parameter is a crucial issue to
be studied because it can possibly establish a relationship
between color deconfinement and strangeness enhancement.

\subsection{Heavy Ions} \label{ssec:hi}

\begin{figure}
\caption{Calculated $\ls$ in the NJL model in a color singlet
cluster with $T = 158$ MeV, baryon chemical potential $\mu_{B}/T =
1.5$, $\mu_Q = 0$ and $\mu_s=0$, as a function of the volume. The
dashed line indicates the grand-canonical value of $\ls$. The constituent
quark masses have been fixed to their values in the
grand-canonical limit: $M_u= M_d= 103$ MeV and $M_s = 452$ MeV.
\label{lshivc}}
\includegraphics{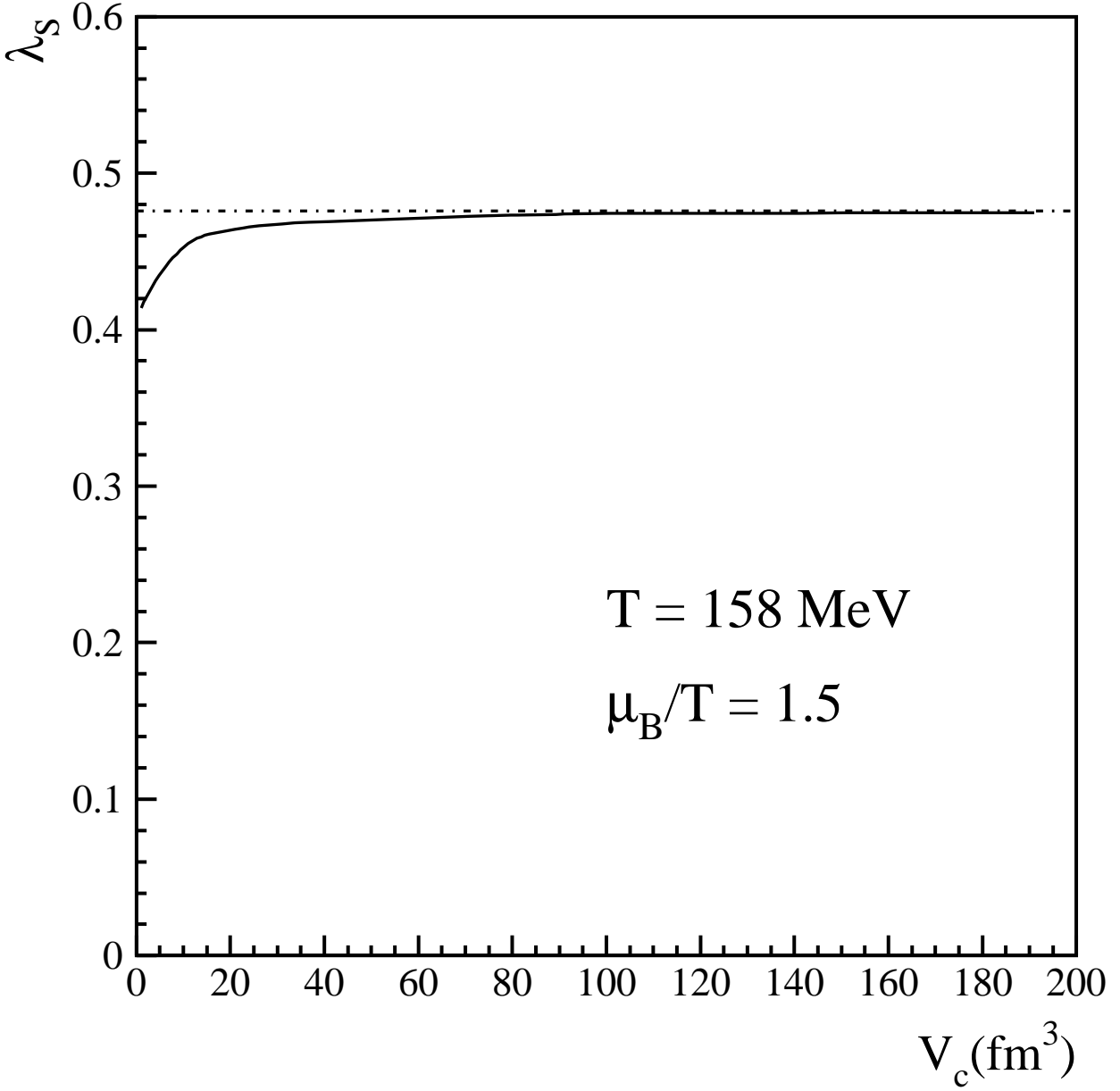}
\end{figure}

In HIC, calculations are much simpler as the total volumes
(namely the sum of proper volumes of all the clusters) turn out
to be so large that the effect of the canonical suppression related
to exact conservation of
flavor quantum numbers can be disregarded \cite{keranen}, provided
that flavor quantum numbers are distributed among clusters
according to the function~(\ref{w}). Indeed, it should be pointed
out that different scenarios have been devised in which the volume
within which strangeness exactly vanishes ({\em strangeness
correlation volume}) is less than the total volume, but this
approach cannot account for the low yield of $\phi$ meson
\cite{redlich}, so we will stick to the identification between the
strangeness correlation volume and the total volume. If $V_c$ is
large enough, quark multiplicities can thus be estimated by means
of Eq.~(\ref{freenumb}), the grand-canonical formula, and
$\lambda_{S}$ turns out to be independent of $V$. In fact, it must
be emphasized that the total volumes fitted in the SHM for HIC,
though affected by large errors, lie in a range (see Table~\ref{tabbeca})
where canonical suppression is negligible and $\lambda_S$ is sensitive
to $V_c$ if this is roughly below 10 fm$^3$ (see Fig.~\ref{lshivc}).
This little sensitivity to $V_c$ makes $\lambda_S$ certainly not
a clear-cut probe for deconfinement, though hadronizing clusters
as small as some fm$^3$ can indeed be excluded.

\begin{figure}
\caption{Cross-over curve in the NJL model with $T_{0}=170$ MeV in
the $(\mu_{B},T)$ plane. Black dots with error bars are the values
obtained within the SHM analysis in various heavy ion reactions
~\cite{becahions2},\cite{becasqm}.
\label{crossline}}
\includegraphics{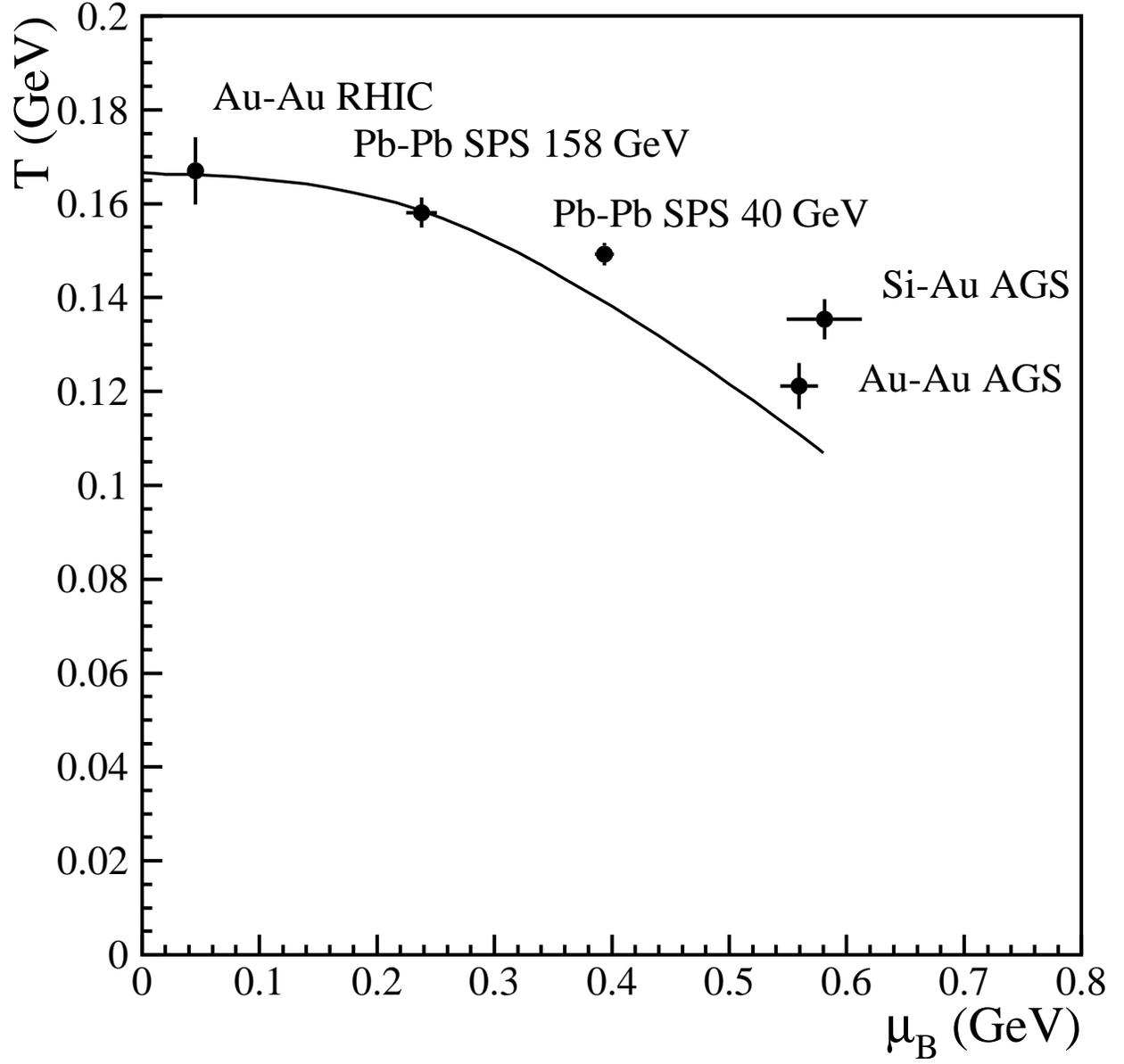}
\end{figure}

Hence, by taking a full grand-canonical approach, the cross-over
line for chiral symmetry breaking has been calculated by
minimizing the light quark mass susceptibility $\chi_m$. As has
been mentioned, by forcing the location of the minimum to coincide
with the fitted $T$ and $\mu_B$ in Pb--Pb collisions at $\sqrt
s_{NN} = 17.3$ GeV, we have been able to set the parameter $T_0$
to 170 MeV (see Fig.~\ref{susc}). For this calculation, the u and
d chemical potentials have been assumed to be equal, so that
$\chi_u$ and $\chi_d$ coincide with $\chi_m$, whereas the strange
quark chemical potential $\mu_s$ has been set to zero due to $S=0$
constraint. It must be noted that, in principle, $\mu_u$ and
$\mu_d$ differ because:

\begin{eqnarray}
  \mu_u &=& \frac{\mu_B}{3} +\frac{2}{3}\mu_Q \nonumber \\
  \mu_d &=& \frac{\mu_B}{3} -\frac{1}{3}\mu_Q \nonumber \\
  \mu_s &=& \frac{\mu_B}{3} -\frac{1}{3}\mu_Q - \mu_S
\end{eqnarray}
where $\mu_Q$ is the electrical chemical potential, which vanishes
only in isospin symmetric systems. However, this term turns out
to be negligible for all examined collisions (see Table~\ref{tabbeca})
and we have simply taken $\mu_u = \mu_d = \mu_B/3$.
Having set $T_0$, the cross-over curve can be predicted from the model
and this is shown in Fig.~\ref{crossline} along with fitted SHM points
which are in satisfactory agreement with the calculation.

It must be pointed out that in a grand-canonical finite system
with fixed volume, the conservation of a definite initial values
of baryon number, electric charge and strangeness necessarily
leads to different values of the chemical potentials for the two
different phases, such as the considered constituent quark phase
and the hadron gas. Certainly, the volume is not supposed to be
fixed in the transition, thus the assumption of equal $\mu_B$'s
may be correct. On the other hand, once $\mu_B$ has been set, the
constraints $S=0$ and $Q/B=Z/A$ lead to definite $\mu_Q$ and
$\mu_S$ (beware the difference with $\mu_s$) values which, in
principle, are different in the two phases. While retaining
$\mu_s=0$, we have checked the accuracy of the assumption
$\mu_Q=0$ by studying the effect on $\ls$ of non-vanishing
electrical chemical potentials in the quark phase. The procedure
is as follows: first, we have calculated $\ls$ and quark masses
with the main assumption, i.e. $\mu_Q=0$; then, by using those
masses, we have calculated $\mu_Q$ in the quark phase by enforcing
$Q/B=Z/A$; finally, with the obtained $\mu_Q$, we have
recalculated quark masses and the new $\ls$. The difference
between the $\ls$'s calculated in the two ways ranges from 1\% in
Pb--Pb collisions at SPS energy up to 2.7\% in Au--Au collisions
at AGS energy and is therefore negligible throughout.

\begin{figure}
\caption{$\ls$ calculated along the cross-over curve predicted in the
NJL model with $T_{0}=170$ MeV. Black dots with error bars are the
values obtained within the SHM analysis in various heavy ion reactions
\cite{becahions2,becasqm}.
\label{lscrit}}
\includegraphics{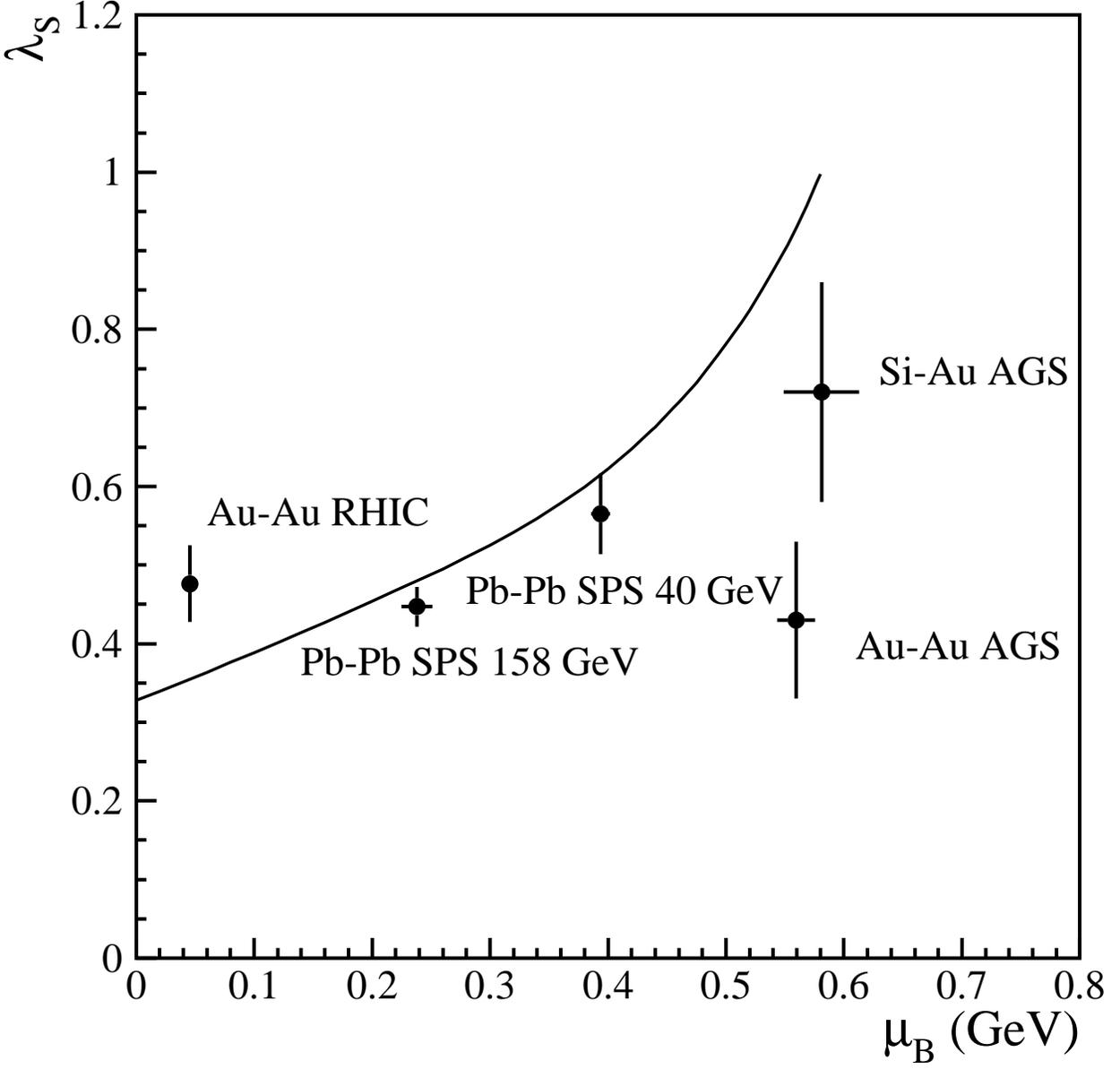}
\end{figure}

The parameter $\lambda_S$ has been calculated along the cross-over
curve by using Eq.~(\ref{freenumb}) and the comparison with the
SHM fitted values is shown in Fig.~\ref{lscrit}. The main
prediction of the NJL-based model is an increase of $\lambda_S$
for decreasing $\sqrt s$ which is driven by the increase of
$\mu_B$ (see Table~\ref{tabbeca}). This involves an enhancement of
relative strange quark production with respect to u, d, the
so-called Pauli blocking effect. On the other hand, the observed
$\lambda_S$ does not keep growing but undergoes a dramatic drop at
very small energies (see Fig.~\ref{lambdas}), a fact which can be
possibly explained by the onset of a purely hadronic production
mechanism at low (yet not easy to locate) center-of-mass energies
with possible local strangeness conservation \cite{redlich}.
Looking at the deviations of the SHM-fitted central values of
$\ls$ in Au--Au and Si--Au from the theoretical curve in
Fig.~(\ref{lscrit}), it can be argued that this hadronic production
mechanism takes over at energies as low as AGS's. Certainly, error
bars are large and a definite conclusion needs more precise data.
It should also be pointed out that a mild strange canonical
suppression sets in already in AGS Si--Au collisions\cite{becahions}
which is not taken into account here. Another
sizeable discrepancy between data and model shows up in Au--Au
collisions at RHIC ($\sqrt{s}=130$ GeV). However, in this case,
$\lambda_S$ has been determined by fitting ratios of hadronic
yields measured at mid-rapidity, unlike all other quoted
collisions for which full phase space multiplicities have been
used. This method may have led to an overestimation of $\ls$ if
the midrapidity region is enriched in strangeness as observed in
Pb--Pb collisions at SPS\cite{becasqm}.

\begin{figure}
\caption{Constituent masses of the quark s and the light quarks u,
d as a function of temperature, with fixed $\mu_{u,d}/T=0.5$ and
$\mu_s =0$. The three curves correspond to different $T_{0}$
values. The vertical lines define the 1$\sigma$ band for Pb-Pb
collisions at $\sqrt{s}_{NN}=17.3$ GeV, as fitted in the SHM
\cite{becahions2}. \label{masses}}
\includegraphics{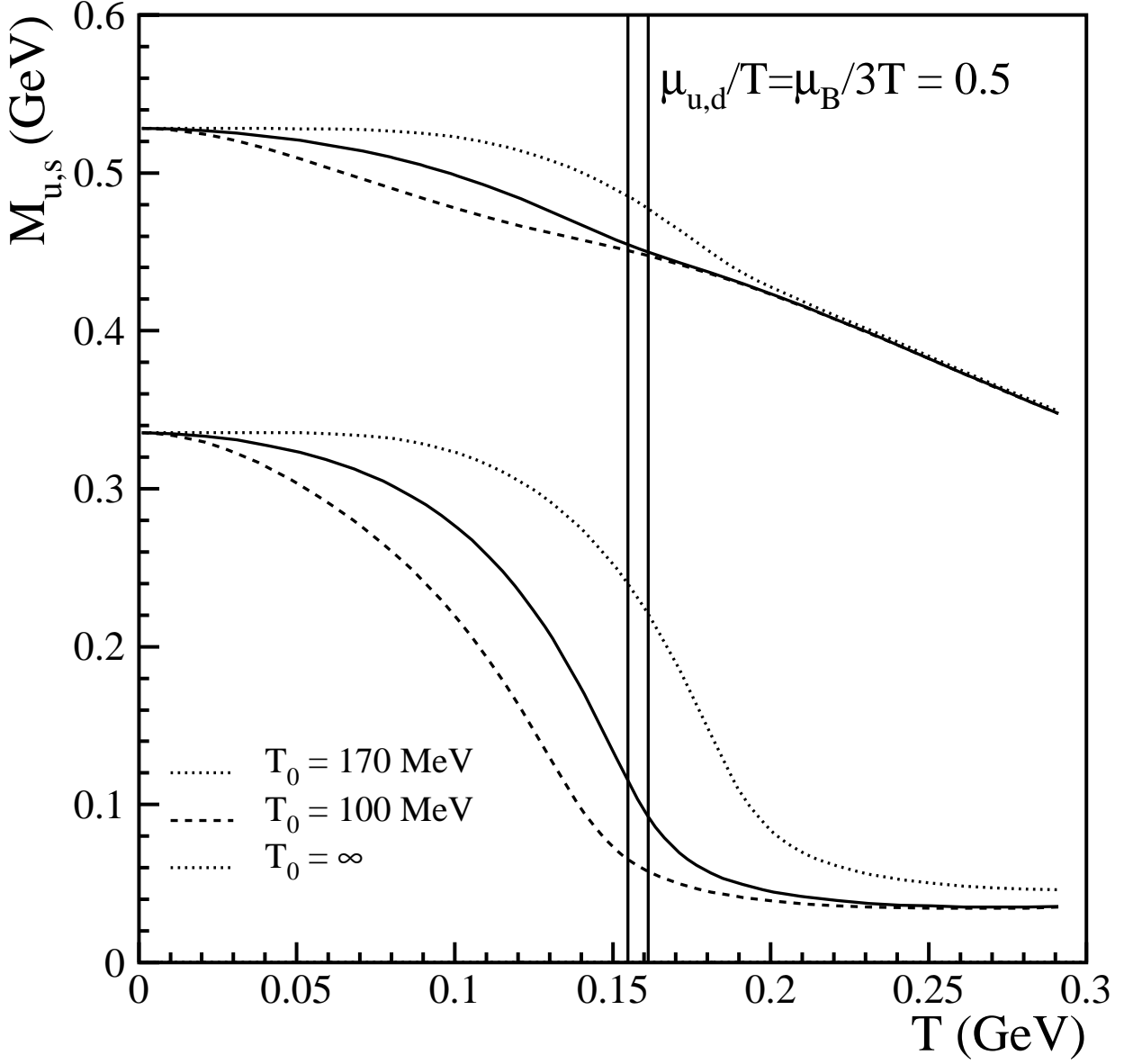}
\end{figure}

\begin{figure}
\caption{Allowed regions of constituent quark masses $M_{u,d},
M_{s}$ determined by the parameters $\ls = 0.447\pm 0.025$,
$T=158.1\pm 2.3$ and $\mu_B=238\pm 13$ MeV as fitted in Pb-Pb
collisions at $\sqrt{s}_{NN}=17.3$ GeV~\cite{becahions2}, with
$\mu_Q=0$ and $\mu_s=0$. The lighter hatched region
corresponds to the $\Lambda$ momentum cut-off as in the NJL model
\cite{kunihiro} whereas the hatched darker region to the free
constituent quark gas (i.e. $\Lambda=\infty$). The solid line
shows the predictions of the full NJL model with $T_{0}$ ranging
from zero (no KMT term) to $\infty$ (CASE I in
ref.~\cite{kunihiro}) from left to right, in the grand-canonical
ensemble with $T=158.1$ MeV and $\mu_B=238$ MeV while the black dot
shows the values of masses for $T_0=170$ MeV. The dash-dotted
horizontal and vertical lines correspond to the $T=\mu_{B}=0$
constituent masses values for the strange quark and for the light
quarks u,d respectively.
\label{bande}}
\includegraphics{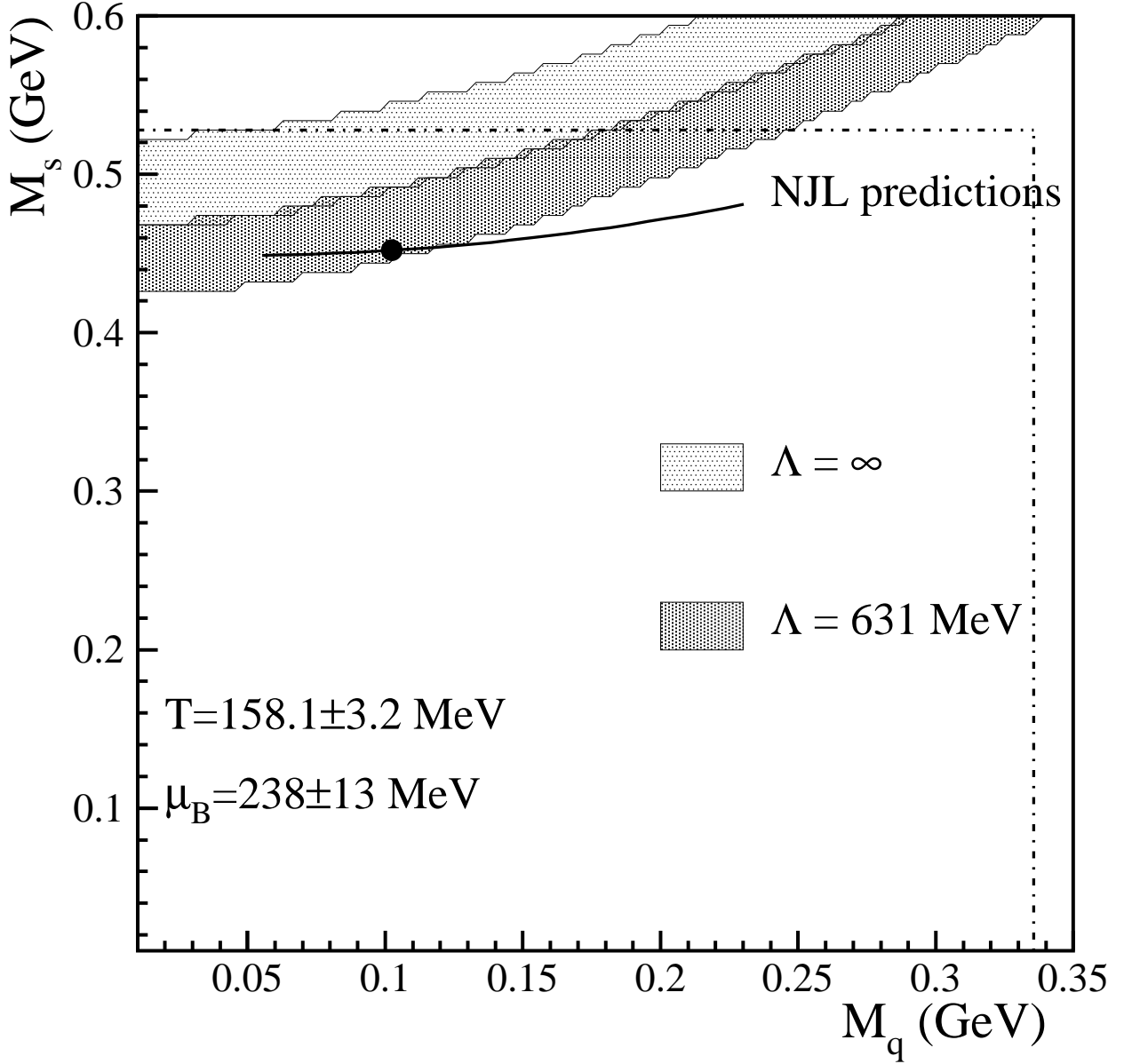}
\end{figure}

It is now worth discussing the robustness of these results in some
detail. The calculation of $\lambda_S$ based on Eq.~(\ref{freenumb})
depends on constituent quark masses and the UV cutoff parameter $\Lambda$
besides the thermodynamical parameters $T$ and $\mu_B$ which have been
set by using hadronic fits. This expression of $\ls$ is indeed a general
one for effective models with four fermion interactions in the mean-field
approximation. It is then worth studying $\ls$ by taking the constituent
masses as free parameters instead of fixing them by means of a particular
effective model as we have done so far. Whilst $M_{u,d}(\mu,T)$ is strongly
dependent on $T_0$, $M_{s}(\mu,T)$ is fairly independent of it
(see Fig.~\ref{masses}). Therefore, as far as the finite temperature
sector is concerned, one can fairly conclude that this NJL model has
actually one free parameter, either $T_0$ or $M_{u,d}(\mu_{Bc},T_c)$.
In other words, fixing $T_0$ amounts to set a definite value of the
light quark mass at the critical point.\\
However, the resulting $\ls$ value does not strongly depend on
this parameter either, as shown in Fig.~\ref{bande}, where the
region in the $(M_u,M_s)$ plane (with both $\Lambda=631.4$ MeV and
$\Lambda=+\infty$) allowed by the fitted $\ls$ values in Pb--Pb
collisions at $\sqrt s_{NN} = 17.3$ GeV is shown: the bands
pattern clearly indicates that $\ls$ has a strong dependence on
$M_s$ and much milder on $M_{u,d}$. Furthermore, the $M_s$ value
at the critical temperature is quite constrained by its
corresponding value at zero temperature and does not undergo
strong variations as a function of the temperature and $T_0$, as
shown in Fig.~(\ref{masses}), which is indeed a quite general
feature of effective models ~\cite{conmassa}. Summarizing, we can
state that this NJL model yields a $\ls$ value in good agreement
with the data (as determined through the SHM) essentially because
it has a fit value of the constituent strange quark mass at zero
temperature and density (about 500 MeV), whilst the particular
value of $T_0$ has much less impact on it (see also
Fig.~(\ref{bande})). On the other hand, for fixed constituent
quark masses, $\ls$ is sizeably affected by the cutoff $\Lambda$.
The calculation with $\Lambda=+\infty$ is meant to give some upper
bound on its variation due to neglected UV contributions, though a
re-analysis within a renormalizable model would be desirable to
have a more accurate estimation.

\subsection{Elementary collisions} \label{ssec:ec}

In EC the fitted total volumes are small and the calculations have to
be carried out within the canonical formalism. As has been mentioned
in the previous section, the partition function to be used is that
in Eq.~(\ref{zetafinal}) with constituent masses replacing current
masses and a $\Lambda$ cut-off as upper bound in the momentum integration
in Eq.~(\ref{totalhatz}). Unlike in the grand-canonical
ensemble, $\ls$ depends on the volume and a thorough comparison with
SHM fitted parameters is much more involved. This is true in many
respects: first, the fitted total volumes in the SHM are subject to
large errors and obtained for point-like hadrons, thus reasonably
underestimated. Secondly, for finite volumes one should in principle
refit $T_0$ so as to obtain a cross-over point for a definite total
volume $V$ and cluster volume $V_c$, for each process, at the desired
temperature. Finally, one should minimize the full expression of
free energy $F = - T \log Z$ (with Z as in~(\ref{zetafinal})) to
determine the constituent masses at finite volume. For the present,
this calculation is not affordable as the minimization of functions
involving five-dimensional numerical integration as in  Eq.~(\ref{zetafinal})
at each step with the due accuracy, implies exceedingly high computing
times with presently available computers. For this reason, we have
carried out a simpler calculation, with $T_0$ kept to 170 MeV and
constituent quark masses calculated in the model at the thermodynamic
limit.

Yet, even in this approximated calculation, the effect of flavor
and color conservation over finite volumes can be studied and,
hopefully, it can be verified if this mechanism of canonical
suppression actually implies a reduction of $\lambda_{S}$ with
respect to the thermodynamic limit consistently with the data. In
fact, any conservation law enforced on finite volumes imply a
reduction of heavy charged particle multiplicity stronger than
lighter particles' so that a decrease of $\ls$ for decreasing
volumes is expected. The ultimate reason of this effect is the
reduced energy expense needed, in a finite system, to compensate
any charge unbalance with light particles in comparison with heavy
particles.

\begin{figure}
\caption{Plot of $\lambda_{S}$ as a function of the total volume
$V$ of the hadronizing clusters, for various (color singlet)
single cluster volumes $V_c$ in an initially completely neutral
system. The constituent quark masses have been set to their values
obtained in the grand-canonical case with $T=160$ MeV and all
chemical potentials vanishing. The dotted line has been obtained
by disregarding the color singlet constraint and keeping only the
flavor constraint, for a given volume. The arrow indicates the
asymptotic value in the grand-canonical limit. The horizontal band is the
region determined by the maximal spread of central values obtained
in the SHM fits in \ee ~\cite{beca2,becapt,becabiele} and in \ppb
collisions \cite{beca2}.
\label{lsepem}}
\includegraphics{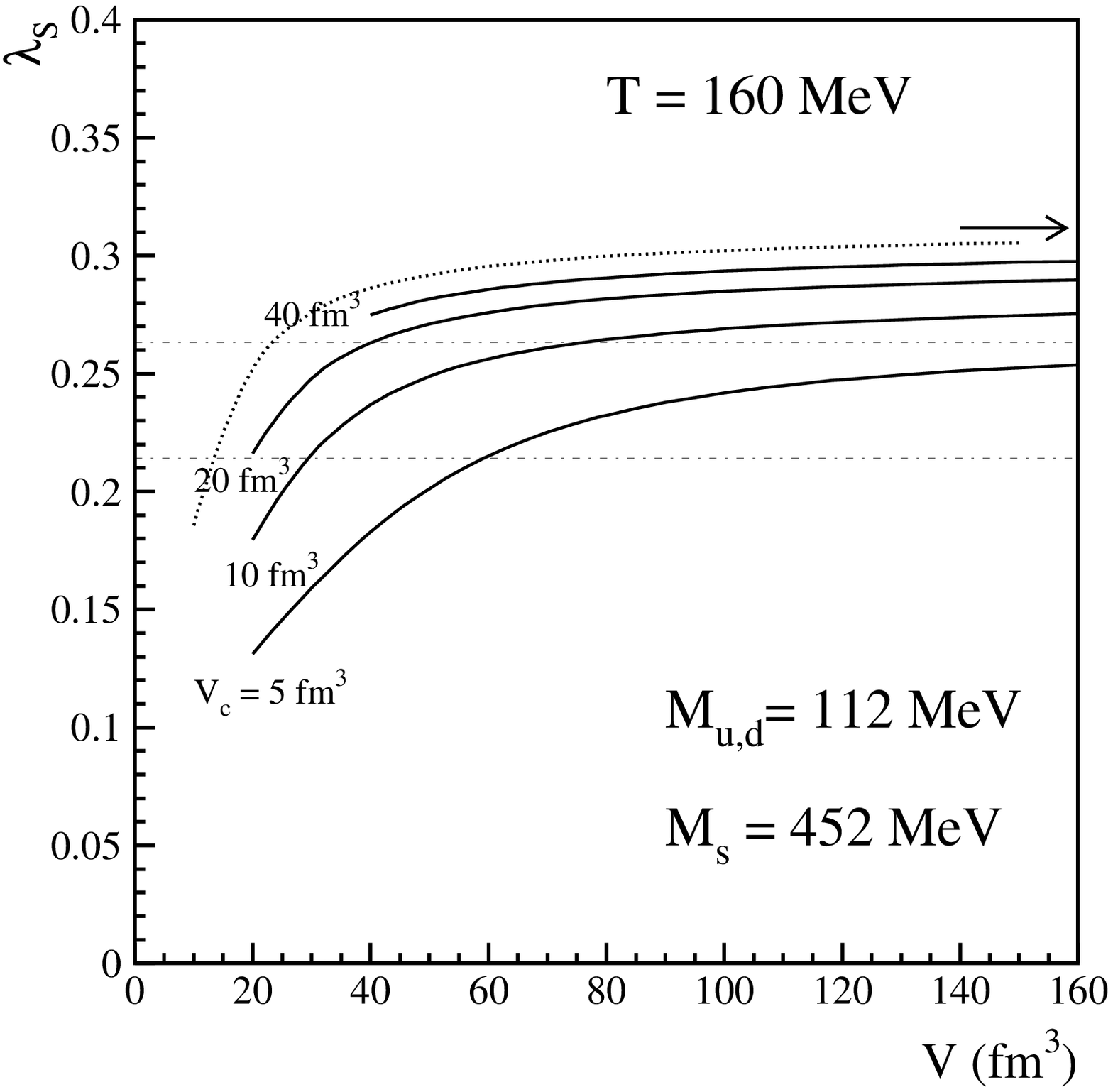}
\end{figure}

We can see this mechanism at play in Fig.~\ref{lsepem} where $\ls$
is plotted as a function of the total volume $V$ (over which
flavor is exactly conserved), for different (color-singlet)
cluster volumes $V_{c}$ for an initially completely neutral
system, such as \ee. The temperature has been set to $T=160$ MeV,
which is a fair average of the SHM fitted values
\cite{beca2,becapt,becabiele} in \ee and the constituent quark
masses have been set to $M_{s}=452$ MeV and $M_{u,d}=112$ MeV,
which are the thermodynamic limit values. Interestingly, $\ls$
features a fair stability over a large range of total volumes
greater than $\approx 20$ fm$^3$ for $V_c < 10$ fm$^3$ within the data
band, with a mild increase up to an asymptotic value which turns
out to be closer to the thermodynamic limit (yet not equal to it,
see also Fig.~(\ref{lshivc})) for larger $V_c$. As the volumes
fitted in a point-like hadron gas model in \ee and \ppb collisions
turn out to be larger than 15 fm$^3$, but not larger than $\simeq
80$ fm$^3$, we can argue that a suitable $V_c$ can account for the
observed values and the stability of $\ls$ over a reasonably large
volume range, taking into account that the actual volumes are
certainly larger than their point-like estimates. Moreover, the
mechanism of local color neutrality
induces the reduction of $\ls$ value with respect to the
thermodynamic limit of $\simeq 0.31$ which is needed to
quantitative reproduce the data. Within this approach, the
difference between $\ls$ in high energy HIC and EC would be mainly
the result of the non vanishing initial baryon density (from
$\simeq 0.31$ to $\simeq 0.45$) and of local color neutrality in
EC over small regions ($5-10$ fm$^3$).

\begin{figure}
\caption{Plot of $\lambda_{S}$ as a function of the total volume
$V$ of the hadronizing clusters, for various (color singlet)
single cluster volumes $V_c$ in an initially pp-like system
($B=Q=2,~S=0$). The constituent quark masses have been set to
their values obtained in the grand-canonical case with $T=160$ MeV
and all chemical potentials vanishing. The dotted line has been
obtained by disregarding the color singlet constraint and keeping
only the flavor constraint, for a given volume. The arrow
indicates the asymptotic value in the grand-canonical limit. The horizontal
band is the region determined by the maximal spread of central
values obtained in the SHM fits in pp collisions \cite{beca2,becapt}.
\label{lspp}}
\includegraphics{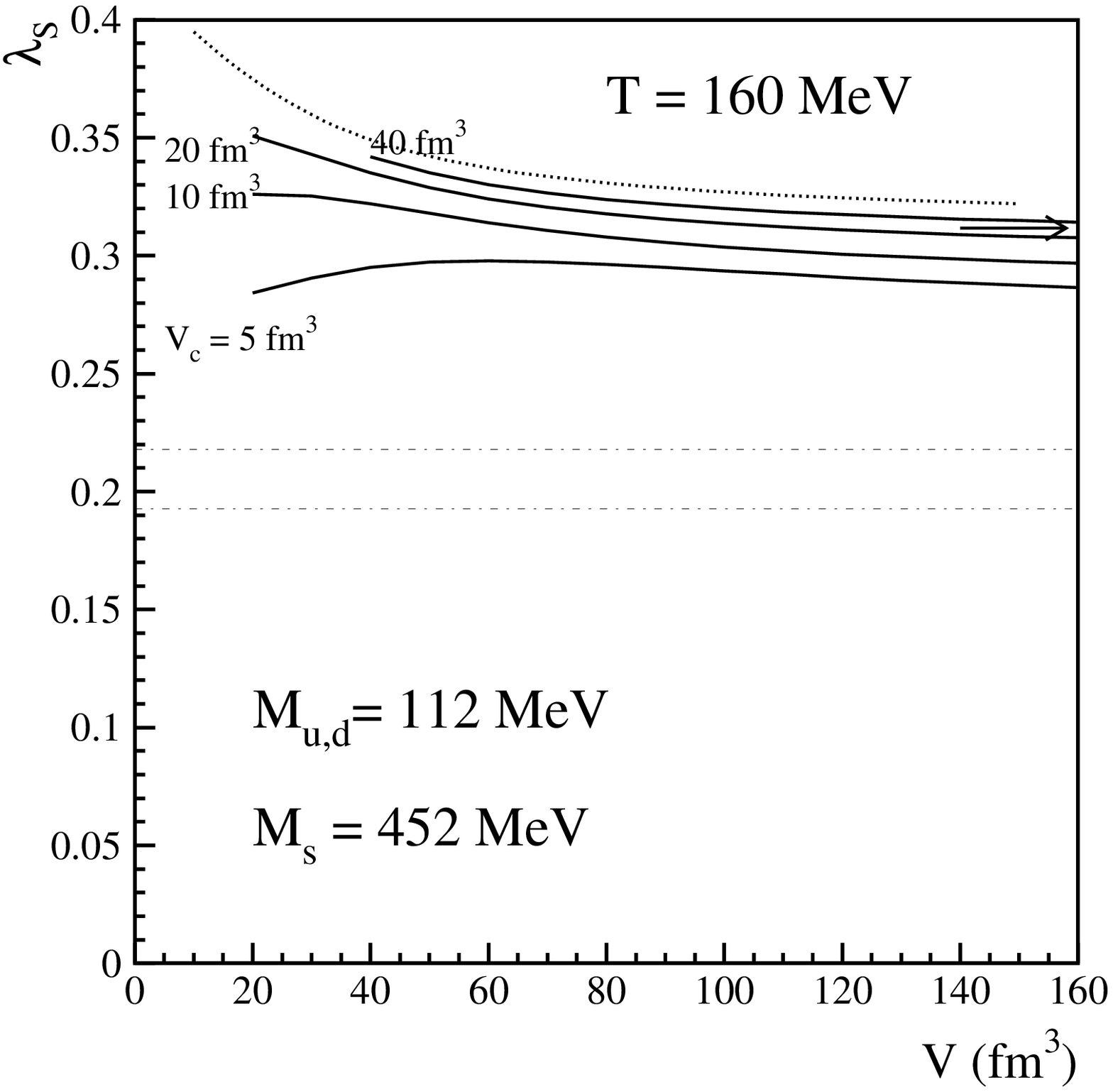}
\end{figure}

The situation is somewhat different for pp collisions, due to non
vanishing initial baryon number and electric charge. In this
system, one can observe two compensating effects, namely Pauli
exclusion principle favoring relative s$\bar{\rm s}$ production and
canonical suppression, favoring relative u$\bar{\rm u}$ and
d$\bar{\rm d}$ production. As a result, $\ls$ is nearly constant
around 0.3, in apparent disagreement with the observed value which
is about 0.2 (see Fig.~\ref{lspp}). The reduction of $\ls$ brought
about by the decrease of $V_c$ is not sufficient to restore the
agreement even for very low $V_c$. The disagreement can be
possibly explained by the inadequacy of the taken assumptions,
such as the statistical distribution of charges among clusters
according to Eq.~(\ref{w}).

\section{CONCLUSIONS AND OUTLOOK}
\label{sec:conclu}

We have presented a statistical model to explain the observed
pattern of strangeness production in both elementary and heavy ion
collisions. The basic idea is that full chemical equilibrium is
locally achieved at the level of constituent quark degrees of
freedom (within the framework of a simple effective model), at a
different stage of the evolution process in elementary collisions
(late) with respect to heavy ion collisions (early). In this
approach, hadron formation takes place through the coalescence of
constituent quarks and this accounts for the known observation of
an incomplete strangeness equilibrium at hadron level in the
statistical hadronization models. The underlying assumption is
that at least the ratio s/u survives in the final hadrons.

Besides the effect of different initial excess of matter over
antimatter, the smaller relative strangeness production in EC with
respect to HIC has been related to the smaller overall system size
and to color confinement over much smaller regions. Particularly,
the existence of a characteristic cluster volume, roughly between
5 and 10 fm$^3$ and independent of center-of-mass energy, can
possibly account for the observed stability of $\ls$, along with
the constancy of temperature and the weak dependence of $\ls$
itself on the total volume $V$ (if not below $\sim$60 fm$^3$), at
least in \ee and \ppb collisions. However, the dependence of $\ls$
on cluster volume in HIC is so mild that it is not possible to use
strangeness production to prove the formation of large
color-neutral regions, i.e. color deconfinement.

The numerical analysis, carried out within an effective
Nambu-Jona-Lasinio model, involved only one parameter to be
adjusted which has been used to get the critical point agreeing
with that fitted in Pb--Pb collisions. A satisfactory agreement
with the data has been found in several heavy ion collisions
and \ee, \ppb collisions. Deviations have been found at low
energy HIC, such as Au--Au and Si--Au at AGS, where the strangeness
production mechanism could be predominantly driven by hadronic
inelastic collisions. Furthermore, a significant discrepancy has
been found in pp, which might be perhaps cured by
taking a different scheme of quantum numbers distribution among
the hadronizing clusters. It should be pointed out that the
analysis in EC has been performed with some approximations as a
full consistent treatment of the Nambu-Jona-Lasinio model for
finite volumes is presently beyond our possibilities.

The model-dependence of our results has been discussed. We are reasonably
confident that the obtained results are quite general and stable
within non-renormalizable models with four-fermion interactions treated
in the mean field approximation. The exploration of renormalizable effective
model is under investigation.

\section*{APPENDIX A}

We want to prove that:

\begin{equation}\label{app1}
\sum_{\Q1,\ldots,\QN} \prod_{i=1}^N Z_i(\Qi) \, \delta_{\Qz,\Sigma_i \Qi}
\end{equation}
is the global partition function in Eq.~(\ref{zetafinal}), with $Z_i$ as
in Eq.~(\ref{zeta}), provided that all clusters have the same volume
$V_c$. We will first prove it in the canonical framework, with each cluster
having also the same temperature $T$. One can rewrite Eq.~(\ref{app1}) by
using Eq.~(\ref{zeta}) and the integral representation of the
$\delta_{\Qz,\Sigma_i \Qi}$ as:

\begin{equation}
\sum_{\Q1,\ldots,\QN} \left[ \prod_{j=1}^3 \int_{-\pi}^{\pi}
{d \phi_j \over 2\pi} \right] e^{\displaystyle{\i \Qz \cdot \phiv -\i
\Sigma_i \Qi \cdot \phiv}} \prod_{i=1}^N \left[ \prod_{j=1}^3
\int_{-\pi}^{\pi} {d \phi_{ji} \over 2\pi} \right] e^{\displaystyle{\i \Qi
\cdot \phiv_i}} \int d \mu(\theta_1,\theta_2) \;
e^{\displaystyle{V_c f(\theta_1,\theta_2,\phiv_i)}}
\end{equation}
where $\hat Z_i$ has been written in the simple form $\exp[V_c f(\theta_1,
\theta_2, \phiv_i)]$ according to Eq.~(\ref{totalhatz}). It must be stressed
that this expression of $\hat Z$ is very general and does not depend on
the Hamiltonian of the considered model. Moreover, the function $f$ is
the same for all clusters for it depends only on the temperature
which has been assumed to be the same. We can now perform the summation
over all $\Qi$ in Eq.~(\ref{app1}) and get:

\begin{equation}
 \sum_{\Q1,\ldots,\QN} e^{\displaystyle{\i \Sigma_i \Qi \cdot
 (\phiv_i -\phiv)}} = \prod_{i=1}^N (2 \pi)^3 \, \delta^3 (\phiv- \phiv_i)
\end{equation}
so that the integration over $\phi_{ji}$ in Eq.~(\ref{app1}) can be easily
performed and one is left with:

\begin{equation}\label{app2}
\left[ \prod_{j=1}^3 \int_{-\pi}^{\pi} {d \phi_j \over 2\pi} \right]
e^{\displaystyle{\i \Qz \cdot \phiv}} \prod_{i=1}^N \int d \mu(\theta_1,\theta_2) \;
e^{\displaystyle{V_c f(\theta_1,\theta_2,\phiv)}}
\end{equation}
The integral over SU(3) group is the same for all clusters and Eq.~(\ref{app2})
can then be written as:

\begin{eqnarray}
&& \left[ \prod_{j=1}^3 \int_{-\pi}^{\pi} {d \phi_j \over 2\pi} \right]
e^{\displaystyle{\i \Qz \cdot \phiv}} \left[ \int d \mu(\theta_1,\theta_2) \;
e^{\displaystyle{V_c f(\theta_1,\theta_2,\phiv)}} \right]^N \nonumber \\
= && \left[ \prod_{j=1}^3 \int_{-\pi}^{\pi} {d \phi_j \over 2\pi} \right]
e^{\displaystyle{\i \Qz \cdot \phiv}} \left[ \int d \mu(\theta_1,\theta_2) \;
{\hat Z}_i (\theta_1,\theta_2,\phiv)\right]^{V/V_c}
\end{eqnarray}
where $V= \Sigma_i V_i = N V_c$, which is equal to the right hand side of
Eq.~(\ref{zetafinal}).

\section*{APPENDIX B}

We show that the configuration probabilities $w(\Q1,\ldots,\QN)$
in Eq.~(\ref{w}) minimize the free energy of a system with volume
$V$ and temperature $T$ which is split into $N$ color-singlet
clusters with volumes $V_1,\ldots,V_N$ such that $\sum_i V_i = V$.
Let $p$ be the full probability of a single state in this system
and $w$ the probability of a configuration $(\Q1,\ldots,\QN)$.
Then:

\begin{equation}
  p_{\rm state} = w(\Q1,\ldots,\QN) \prod_{i=1}^N \frac{\exp(-E_i/T)}{Z_i(\Qi)}
  \delta_{\Qi,\Qis} \delta_{\rm singlet_i}
\end{equation}
where $Z_i$ is given by Eq.~(\ref{zeta}), $E_i$ is the energy of a
cluster, $\Qis$ is the vector of flavor quantum numbers of the
state and $\delta_{\rm singlet_i}$ signifies the color singlet
constraint on each cluster. The entropy reads:

\begin{eqnarray}\label{entropy}
 S &=& - \sum_{\rm states} p \log p \nonumber \\
 &=& - \!\!\!\!\!\sum_{\Q1,\ldots,\QN} \sum_{\rm states_1} \!\!\! \ldots \!\!\!
 \sum_{\rm states_N}
 \!\!\! w(\Q1,\ldots,\QN) \prod_{i=1}^N \frac{e^{\displaystyle{-E_i/T}}}{Z_i(\Qi)}
 \log w (\Q1,\ldots,\QN) \left[ \prod_{i=1}^N \frac{\exp(-E_i/T)}{Z_i(\Qi)} \right]
 \nonumber \\
 &&
\end{eqnarray}
where the flavor and color constraint on each cluster are now
implied in the sum over the states.

Eq.~(\ref{entropy}) can be worked out:
\begin{eqnarray}
 S & = & - \sum_{\Q1,\ldots,\QN} \sum_{\rm states_1} \ldots \sum_{\rm states_N}
 w(\Q1,\ldots,\QN) \prod_{i=1}^N \frac{e^{\displaystyle{-E_i/T}}}{Z_i(\Qi)} \nonumber \\
 & \times & \left[ \log w (\Q1,\ldots,\QN) - \frac{E_i}{T} -
 \log \prod_{i=1}^N Z_i(\Qi) \right] \nonumber \\
 & = & \!\!\!\!\! \sum_{\Q1,\ldots,\QN} \!\!\! w(\Q1,\ldots,\QN)
    \left[ -\log w(\Q1,\ldots,\QN)
  + \frac{\sum_{i=1}^N \langle E_i \rangle_{(\Q1,\ldots,\QN)}}{T} +
  \log \prod_{i=1}^N Z_i(\Qi) \right] \nonumber \\
 &&
\end{eqnarray}
where $\sum_i \langle E_i \rangle_{(\Q1,\ldots,\QN)}$ is the average
energy for the $i^{\rm th}$ cluster and a fixed configuration $(\Q1,\ldots,\QN)$.
The total average energy of the system, to be identified with the internal energy
$U$, evidently reads:

\begin{equation}
 U = \sum_{\Q1,\ldots,\QN} w(\Q1,\ldots,\QN) \sum_{i=1}^N
 \langle E_i \rangle_{(\Q1,\ldots,\QN)}
\end{equation}
so that entropy can be rewritten as:

\begin{equation}
 S = \!\!\! \sum_{\Q1,\ldots,\QN} -w(\Q1,\ldots,\QN) \log w(\Q1,\ldots,\QN) +
    \frac{U}{T} + \!\!\! \sum_{\Q1,\ldots,\QN} \!\!\! w(\Q1,\ldots,\QN)
    \log \prod_{i=1}^N Z_i(\Qi)
\end{equation}
and the free energy $F$ as:

\begin{equation}
 F = U - TS = T \sum_{\Q1,\ldots,\QN} w(\Q1,\ldots,\QN) \left[ \log w(\Q1,\ldots,\QN) -
     \log \prod_{i=1}^N Z_i(\Qi) \right]
\end{equation}
Now we seek for the probabilities $w$ which minimize the free
energy with the constraint $\sum w = 1$. Therefore we have to
introduce a Lagrange multiplier $\lambda$ and look for the
extremals of $F + \lambda (\sum w - 1)$ with respect to each
$w(\Q1,\ldots,\QN)$ and $\lambda$. This leads to:

\begin{equation}
  \log w(\Q1,\ldots,\QN) = \log \prod_{i=1}^N Z_i(\Qi) - \frac{\lambda}{T}
\end{equation}

and
\begin{equation}
  w(\Q1,\ldots,\QN) = \frac{\prod_{i=1}^N Z_{i}(\Qi)
  \delta_{\Qz,\Sigma_i \Qi}}{\sum_{\Q1,\ldots,\QN} \prod_{i=1}^N
  Z_{i}(\Qi) \delta_{\Qz,\Sigma_i \Qi}}
\end{equation}
after proper normalization.

\begin{acknowledgments}
We are grateful to J. Aichelin, A. Barducci, R. Casalbuoni, U.
Heinz, J. Lenaghan, O. Scavenius for useful discussions.
\end{acknowledgments}



\newpage

\baselineskip = 2\baselineskip  

\begin{table}
\caption{Parameters determined through statistical model fits to
measured multiplicities in some elementary and heavy ion collisions.
Heavy ion parameters have been taken from refs.~\cite{becahions2,becasqm}
whilst elementary collisions from refs.~\cite{beca2,becapt}; \ee at
$\sqrt s=$ 91.2 and \ppb points have been refitted with updated data
and hadronic input parameters. The pp point has been fitted with a
different parametrization of the strangeness suppression \cite{becapt}
and $\gamma_S$ has been replaced with the mean value of strange quark
pairs $\ssb$; the corresponding $\gamma_S$ would be $\simeq 0.5$.
The numerical values in Pb-Pb at
$\sqrt s = 8.7$ GeV have been obtained in ref.~\cite{becasqm} with
preliminary data from experiment NA49, those in Au-Au collisions at
$\sqrt s = 130$ GeV from a numerical analysis of midrapidity ratios
as quoted in ref.~\cite{pbm}. The quoted central values of volumes
(the volume is defined as the sum of the volumes of all produced
clusters), have been obtained with pointlike hadrons, thus they must
be taken as lower limits for the actual values. The errors on volumes,
not quoted here, are generally large and can be up to 50\%. Also quoted the
central values of the electrical chemical potentials obtained in heavy
ion collisions.}
\begin{tabular}{cccccccc}

 Collision &$\sqrt s$ (GeV)& $T$ (MeV)     &$\mu_B$ (MeV) & $\gs$      &  $\ls$     & $\mu_Q$ (MeV)& $V$ (fm$^3$) \\
\tableline
   Au-Au   &      130      & 167$\pm$7.2   & 45.8$\pm$6.4 &1.04$\pm$0.10   &0.476$\pm$0.049 & -1.42        &  -       \\
   Pb-Pb   &     17.3      & 158.1$\pm$3.2 & 238$\pm$13   &0.789$\pm$0.0582&0.447$\pm$0.025 & -6.87        & 3460     \\
   Pb-Pb   &      8.7      & 149.0$\pm$2.4 & 393.7$\pm$8.3&0.822$\pm$0.058 &0.585$\pm$0.052 & -11.2        & 2067     \\
   Si-Au   &      5.4      & 133.4$\pm$4.3 & 581$\pm$32   &0.845$\pm$0.101 &0.72$\pm$0.14   & -10.7        & 330      \\
   Au-Au   &      4.8      & 121.2$\pm$4.9 & 559$\pm$16   &0.697$\pm$0.091 &0.43$\pm$0.10   & -12.4        & 2805     \\
\tableline
   \ee     &     14        & 167.4$\pm$6.5 & & 0.795$\pm$0.088        & 0.243$\pm$0.036 & & 15.9 \\
   \ee     &     91        & 159.2$\pm$0.8 & & 0.664$\pm$0.014        & 0.225$\pm$0.004 & & 52.4 \\
    pp     &     27.4      & 162.4$\pm$1.6 & & $\ssb$ 0.653$\pm$0.017 & 0.201$\pm$0.005 & & 25.5 \\
   \ppb    &     200       & 175$\pm$11    & & 0.491$\pm$0.056        & 0.214$\pm$0.025 & & 35.5 \\
   \ppb    &     900       & 167.$\pm$9.0  & & 0.533$\pm$0.054        & 0.230$\pm$0.033 & & 77.3 \\
\end{tabular}
\label{tabbeca}
\end{table}

\newpage

\newpage

\newpage

\newpage

\newpage

\newpage

\newpage

\newpage

\newpage

\begin{references}
\bibitem{beca1}
     F. Becattini, Z.Phys. {\bf C69}, (1996) 485.
\bibitem{beca2}
     F. Becattini, in Proc. of ``Universality features of multihadronic
     production and the leading effect", October 19-25 1996, World Scientific
     p. 74, hep-ph 9701275; F.Becattini and U.Heinz, Z.Phys. {\bf C76}, (1997) 269.
\bibitem{fermi}
     E. Fermi, Progr. Theor. Phys. {\bf 5} (1950) 570.
\bibitem{hagedorn}
     R. Hagedorn, N. Cim. Suppl. {\bf 3} (1965) 147.
\bibitem{heinz}
     U. Heinz, Nucl. Phys. {\bf A661} (1999) 140.
\bibitem{stock}
     R. Stock, Phys. Lett. {\bf B456} (1999) 277.
\bibitem{hagedorn2}
     R. Hagedorn, CERN lectures ``Thermodynamics of strong interactions"
     (1970) 46.
\bibitem{hions}
     see for instance: P. Braun-Munzinger et al., Phys. Lett. {\bf B344} (1995) 43;
     P. Braun-Munzinger et al., Phys. Lett. {\bf B365} (1996) 1;
     J. Cleymans et al., Z. Phys. {\bf C74} (1997) 319;
     J. Cleymans et al., Phys. Rev. {\bf C59} (1999) 1663;
     W. Florkowski et al., Acta Phys. Polon. {\bf B33} (2002) 761.
\bibitem{becahions2}
     F. Becattini, J. Cleymans, A. Keranen, E. Suhonen, K. Redlich,
     Phys. Rev. {\bf C64} 024901.
\bibitem{becahions}
     F. Becattini, M. Gazdzicki and J. Sollfrank, Eur. Phys. J.
     {\bf C5} (1998) 143.
\bibitem{kinetic}
     S. Bass, A. Dumitru et al., Phys. Rev. {\bf C60} (1999) 021902.
\bibitem{becapt}
     F. Becattini, G. Passaleva, Eur. Phys. J. {\bf C23} (2002) 551.
\bibitem{aichreb}
     P. Rehberg and J. Aichelin, Phys. Rev. {\bf C60} (1999) 064905.
\bibitem{dumitru}
     O. Scavenius, A. Dumitru, J. Lenaghan, hep-ph/0201079.
     Phys. Rev. {\bf D38}, (1988) 238.
\bibitem{heuristic}
     A. Barducci, R. Casalbuoni, S. De Curtis, R. Gatto and G. Pettini,
     Phys. Lett. {\bf B244}, (1990) 311.
\bibitem{karsch}
     F. Karsch and E. Laermann,
     Phys. Rev. {\bf D59} (1999) 031501; F. Karsch hep-lat/9903031.
\bibitem{becabiele} For a review see: F. Becattini,
     ``Statistical hadronisation phenomenology", talk given in
     Statistical QCD, Bielefeld, Aug. 25-30 2001.
\bibitem{njlvero}
     Y. Nambu, G. Jona-Lasinio, Phys. Rev. {\bf 122}, (1991) 345 and
     Phys. Rev. 124, (1991) 246.
\bibitem{kunihiro}
      T. Hatsuda and T. Kunihiro, Phys. Rep. {\bf 247}, (1994) 221.
\bibitem{bcd}
     A. Barducci, R. Casalbuoni, S. De Curtis, D. Dominici and R. Gatto,
\bibitem{tricritical}
     A. Barducci, R. Casalbuoni, S. De Curtis, R. Gatto and G. Pettini,
     Phys. Lett. {\bf B231}, (1989) 463; Phys. Rev. {\bf D41},
     (1990) 1610; Phys. Rev. {\bf D46}, (1992) 2203.
\bibitem{conmassa}
     A. Barducci, R. Casalbuoni, R. Gatto and G. Pettini, Phys.Rev.
     {\bf D49}, (1994) 426.
\bibitem{giap}
     O. Kiriyama, M. Maruyama and F. Takagi, Phys.Rev. {\bf D63}
     (2001) 116009.
\bibitem{fodor} Z. Fodor, hep-lat 0209101 and references therein.
\bibitem{klevalast}
     T.M. Schwarz, S.P. Klevansky  and G. Papp Phys.Rev.{\bf C60}
     (1999) 055205.
\bibitem{aichelin}
     see for instance: F. Gastineau, R. Nebauer and J. Aichelin,
     Phys. Rev. {\bf C65},(2002) 045204.
\bibitem{cfl}
     G. Alford, K. Rajagopal and F. Wilczek, Nucl.Phys. {\bf B422}
     (1998) 247 and Nucl.Phys. {\bf B537} (1999) 443;
     J. Berges, K. Rajagopal, Nucl.Phys. {\bf B538} (1999) 215.
\bibitem{rafe}
     J. Rafelski and B. Muller, Phys. Rev. Lett. {\bf 48} (1982) 1066.
\bibitem{exact}
     K. Redlich and L. Turko, Z.Phys. {\bf C5}, (1980) 201;
     L. Turko,Phys.Lett. {\bf 104B}, (1981) 153; M.I. Gorenstein,
     S.I. Lipskikh, V.K. Petrov and G.M. Zinovjev, Phys. Lett. 123B,
     (1983) 437; M.I. Gorenstein, O.A. Mogilevsky, V.K. Petrov and
     G.M. Zinovjev, Z. Phys. {\bf C18}, (1983) 13; G. Auberson, L. Epele,
     G. Mahoux and F.R.A. Simao, J.Math.Phys. {\bf 27} (6),
     (1986) 1658.
\bibitem{uauno}
     M. Kobayashi and T. Maskawa, Prog.Theor.Phys. {\bf 44}, (1970) 1422;
     M. Kobayashi, H. Kondo and T. Maskawa, Prog.Theor.Phys.
     {\bf 45}, (1971) 1955; G. 'tHooft, Phys.Rev. {\bf D14},
     (1976) 3432; Phys.Rep. {\bf 142}, (1986) 357; M.A. Shifman,
     A.I. Vainshtein and V.Z. Zakharov, Nucl.Phys. {\bf 163B}, (1980) 46.
\bibitem{cjt}
     R. Jackiw, Phys.ReV. {\bf D9}, (1974) 1686; J.M. Cornwall,
     R. Jackiw and E. Tomboulis, Phys.ReV. {\bf D10}, (1974) 2428.
\bibitem{dolan}
     L. Dolan and R. Jackiw, Phys. Rev. {\bf D9}, (1974) 3320;
     C.W. Bernard, Phys.Rev. {\bf D9}, (1974) 3312.
\bibitem{russi}
     see for instance: V.A. Miransky, ``Dynamical Symmetry
     Breaking In Quantum Field Theories", World Scientific (1993),
     and references therein.
\bibitem{keranen}
     A. Keranen, F. Becattini, Phys. Rev. {\bf C65} (2002) 044901.
\bibitem{redlich}
     S. Hamieh, K. Redlich and A. Tounsi, Phys. Lett. {\bf B486} (2000) 61.
\bibitem{becasqm} F. Becattini, talk given at ``Strange Quarks in
  Matter 2001", J. Phys. {\bf G28} (2002) 1553.
\bibitem{pbm} P. Braun-Munzinger et al., Phys. Lett. {\bf B518}
(2001) 41.

\end{references}
\end{document}